\documentclass[aps,preprint,pre,superscriptaddress,runinaddress,showpacs]{revtex4}
\usepackage{amssymb}
\usepackage{amsmath}
\usepackage{graphicx}
\usepackage{bm}

\input{tcilatex}
\begin{document}

\title{Density Functional Theory for the Electron Gas and for Jellium.}
\author{J. W. Dufty}
\affiliation{Department of Physics, University of Florida, Gainesville, FL 32611}
\date{\today }

\begin{abstract}
Density Functional Theory relies on universal functionals characteristic of
a given system. Those functionals in general are different for the electron
gas and for jellium (electron gas with uniform background). However, jellium
is frequently used to construct approximate functionals for the electron gas
(e.g., local density approximation, gradient expansions). The precise
relationship of the exact functionals for the two systems is addressed here.
In particular, it is shown that the exchange - correlation functionals for
the inhomogeneous electron gas and inhomogeneous jellium are the same. This
justifies theoretical and quantum Monte Carlo simulation studies of jellium
to guide the construction of functionals for the electron gas. Related
issues of the thermodynamic limit are noted as well.
\end{abstract}

\maketitle

\section{Introduction}

\label{sec1}Keith Gubbins has been an inspiration to students and colleagues
for more than fifty years, through direct mentoring and collaborations, and
indirectly through his extensive innovative publications. My own
collaboration was very early in my career as a new faculty member in the
Physics Department at the University of Florida. Keith had just finished his
book with Tim Reed showing the relevance of statistical mechanics to
applications in chemical and other engineering fields. He continued to
demonstrate the importance of formulating practical approximations to
complex systems through their firm foundation in basic statistical
mechanics. He also demonstrated by example the need to ignore the artificial
interfaces between applied and basic disciplines, taking recent developments
in the mathematics and physics literature to formulate innovative and
accurate descriptions of complex systems, via the reinforcement of theory,
simulation, and experiment. An example is his early application of non-local
classical density functional theory (DFT) for adsorption in carbon slit
pores, verified by simulation and including a discussion of problems of
extracting experimental parameters needed for the calculation \cite%
{Gubbins97}. The presentation here has in common the tool of density
functional theory for application to a complex system (warm dense matter).
But the system is quite different - confined electrons - and hence quantum
effects can dominate except at the highest temperatures. The emphasis is on
formal relationships to guide practical applications, so it is hoped that
Keith will appreciate some of his spirit in the following.

The historical development of DFT has followed two quite different paths,
one for quantum systems at zero temperature (e.g., electrons in atoms,
molecules, and solids) \cite{Hohenberg64}, \cite{ParrWang89}, \cite%
{Dreizler90}, \cite{Jones15}, and one for temperature dependent classical
non-uniform fluids (e.g., two phase liquids, porous media) \cite{Evans92},  
\cite{Lutsko10}. The former has focused on determination of the ground state
energy whereas the latter has focused on the classical free energy of
thermodynamics. Their relationship was established by Mermin who extended
the original ground state theorems of Hohenberg and Kohn to finite
temperatures for quantum systems \cite{Mermin65}, \cite{Eschrig10} described
by the Gibbs ensemble of statistical mechanics. This early history is
briefly reviewed in references \cite{Jones15} ,\cite{Evans92}. Recently the
two paths have rejoined in the effort to describe conditions of warm, dense,
matter \cite{WDM}. These are solid density conditions but temperatures for
electrons ranging from zero temperature ground state to well above the Fermi
temperature. They include domains with atomic and molecular coexistence, and
with association and dissociation chemistry relevant. A central ingredient
is the (intrinsic) free energy functional $F$, consisting of a
non-interacting contribution $F_{0}$, a Hartree contribution, $F_{H}$, and a
remainder called the exchange - correlation free energy functional, $F_{xc}$%
. The exact functional form for the Hartree contribution is known and the
contribution from $F_{0}$ is treated exactly (numerically) within the Kohn -
Sham formulation of DFT \cite{KS65} extended to finite temperatures. Hence
the primary challenge for applications is construction of the exchange -
correlation density functional.

An important constraint is its equivalence with the corresponding functional
for jellium (electrons in a uniform neutralizing background \cite{Vignale05}%
) when evaluated at a uniform density. Although still a difficult quantity
to determine, the latter has been studied widely by approximate theoretical
methods \cite{Vignale05}, \cite{Sjostrom13} and more recently by quantum
simulation methods across the temperature - density plane \cite{Dornheim16}.
An accurate fitting function for practical applications now exists \cite%
{KSDT15}. Its utility for DFT is within the "local density approximation",
first proposed in reference \cite{KS65} for the zero temperature energy
functional and extended to finite temperatures for the free energy
functional. It assumes that the non-uniform system $F_{xc}$ can be
represented at each point by the uniform jellium $F_{xc}$ evaluated at the
density for that point.

The objective here is to clarify the precise relationship of the DFT
functionals for the two different systems, electron gas and jellium, for
general non-uniform densities. The primary new result is that the exchange -
correlation functionals for the two systems are equivalent, and the total
free energies (intrinsic plus external potential) are the same. In the
analysis it noted that jellium is usually considered in the thermodynamic
limit, $\overline{N}\rightarrow \infty ,$ $V\rightarrow \infty ,$ $\overline{%
N}/V$ where $\overline{N}$ and $V$ are the average particle number and
volume, respectively. It has been proved that the jellium free energy is
well-defined in this limit \cite{Lieb75}, whereas that for the electron gas
is not (it does not scale as the volume for large system size due in part to
lack of charge neutrality). Nevertheless, the exchange - correlation
component of the free energy for the electrons does have a proper
thermodynamic limit as a consequence of the equivalence demonstrated here.
In the next section, Hamiltonians for the isolated electron gas and jellium
are defined. Next, the statistical mechanical basis for DFT is described for
the grand canonical ensemble. The grand potential (proportional to the
pressure) is defined as a functional of a given external potential $%
v_{ex}\left( \mathbf{r}\right) $ (occuring through a local chemical
potential $\mu (\mathbf{r})=\mu -v_{ex}\left( \mathbf{r}\right) $), and the
corresponding functionals for the inhomogeneous electron gas and
inhomogeneous jellium are related. Associated with $\mu (\mathbf{r})$ are
the densities $n_{e}(\mathbf{r})$ and $n_{j}(\mathbf{r})$ for the two
systems. Their relationship is established. Next, the strict concavity of
the grand potential functionals assures a one - to - one relationship of the
densities to $\mu (\mathbf{r})$ so that a change of variables is possible.
This is accomplished by Legendre transformations which define the free
energy density functionals. It is noted that the free energy density
functional obtained in this way is precisely that of Mermin's DFT. The free
energies differ by the potential energy of the background charge. It is
shown that this cancels the differences between the orresponding Hartree
free energy contributions, resulting in the desired equivalence of the
exchange - correlation functionals.

\section{Inhomogeneous electron and jellium functionals}

\label{sec2}The Hamiltonian for an isolated system of $N$ electrons in a
volume $V$ \ is given by%
\begin{equation}
\widehat{H}_{e}=\sum_{\alpha =1}^{N}\frac{\widehat{p}_{\alpha }^{2}}{2m}+%
\frac{1}{2}e^{2}\int d\mathbf{r}d\mathbf{r}^{\prime }\frac{\widehat{n}(%
\mathbf{r})\widehat{n}(\mathbf{r}^{\prime })-\widehat{n}(\mathbf{r})\delta
\left( \mathbf{r}-\mathbf{r}^{\prime }\right) }{\left\vert \mathbf{r}-%
\mathbf{r}^{\prime }\right\vert }.  \label{2.1}
\end{equation}%
where the number density operator is 
\begin{equation}
\widehat{n}(\mathbf{r})=\sum_{\alpha =1}^{N}\delta \left( \mathbf{r-}%
\widehat{\mathbf{q}}_{\alpha }\right) .  \label{2.2}
\end{equation}%
The position and momentum operators for electron $\alpha $ are $\widehat{%
\mathbf{q}}_{\alpha }$ and $\widehat{\mathbf{p}}_{\alpha }$, respectively. A
caret over a symbol denotes the operator corresponding to that variable. The
related jellium Hamiltonian is%
\begin{equation}
\widehat{H}_{j}=\sum_{\alpha =1}^{N}\frac{\widehat{p}_{\alpha }^{2}}{2m}+%
\frac{1}{2}e^{2}\int d\mathbf{r}d\mathbf{r}^{\prime }\frac{\left( \widehat{n}%
(\mathbf{r})-n_{b}\right) \left( \widehat{n}(\mathbf{r}^{\prime
})-n_{b}\right) -\widehat{n}(\mathbf{r})\delta \left( \mathbf{r}-\mathbf{r}%
^{\prime }\right) }{\left\vert \mathbf{r}-\mathbf{r}^{\prime }\right\vert }.
\label{2.3}
\end{equation}%
The constant $n_{b}$ denotes the density of a uniform neutralizing
background for the electrons. When the grand canonical ensemble is
considered it is given by $n_{b}=\overline{N}/V$, where $\overline{N}$ is
the average particle number. The two Hamiltonians are seen to be related by%
\begin{equation}
\widehat{H}_{j}=\widehat{H}_{e}+\int d\mathbf{r}v_{b}\left( \mathbf{r}%
\right) \widehat{n}(\mathbf{r})+E_{b}  \label{2.5}
\end{equation}%
with 
\begin{equation}
v_{b}\left( \mathbf{r}\right) =-e^{2}\int d\mathbf{r}^{\prime }\frac{n_{b}}{%
\left\vert \mathbf{r}-\mathbf{r}^{\prime }\right\vert }.  \label{2.6}
\end{equation}%
The second term on the right side of (\ref{2.5}) is the potential of
interaction between the electrons and the background, and the third term is
the background self energy%
\begin{equation}
E_{b}=\frac{1}{2}\int d\mathbf{r}d\mathbf{r}^{\prime }\frac{\left(
n_{b}e\right) ^{2}}{\left\vert \mathbf{r}-\mathbf{r}^{\prime }\right\vert }=-%
\frac{1}{2}\int d\mathbf{r}n_{b}v_{b}\left( \mathbf{r}\right) .  \label{2.7}
\end{equation}%
Equation (\ref{2.3}) is the usual definition of jellium as the electron
system plus a uniform neutralizing background.

\subsection{Grand potential functionals}

Now consider the addition of an external single particle potential $%
v_{ex}\left( \mathbf{r}\right) $ to the electron and jellium Hamiltonians.
The equilibrium properties for the corresponding inhomogeneous systems are
defined by the grand canonical potentials%
\begin{equation}
\beta \Omega _{e}\left( \beta ,V\mid \mu \right) =-\ln \sum_{N=0}^{\infty
}Tr^{(N)}e^{-\beta \left( \widehat{H}_{e}-\int d\mathbf{r}\mu (\mathbf{r})%
\widehat{n}(\mathbf{r})\right) },  \label{2.8}
\end{equation}%
\begin{equation}
\beta \Omega _{j}\left( \beta ,V\mid \mu \right) =-\ln \sum_{N=0}^{\infty
}Tr^{(N)}e^{-\beta \left( \widehat{H}_{j}-\int d\mathbf{r}\mu (\mathbf{r})%
\widehat{n}(\mathbf{r})\right) },  \label{2.9}
\end{equation}%
Here, the local chemical potential is defined by%
\begin{equation}
\mu (\mathbf{r})=\mu -v_{ex}\left( \mathbf{r}\right) .  \label{2.10}
\end{equation}%
These grand potentials are functions of the inverse temperature $\beta $ and
the volume $V$, and \textit{functionals} of the local chemical potential $%
\mu (\mathbf{r})$. The functionals themselves, $\Omega _{e}\left[ \beta
,V\mid \cdot \right] $ and $\Omega _{j}\left[ \beta ,V\mid \cdot \right] $,
are characterized by $\widehat{H}_{e}$ and $\widehat{H}_{j}$ respectively.
Since the latter two are different, the functionals are different. However,
from (\ref{2.5}) they have the simple relationship%
\begin{equation}
\beta \Omega _{j}\left( \beta ,V\mid \mu \right) =\beta E_{b}+\beta \Omega
_{e}\left( \beta ,V\mid \mu -v_{b}\right) .  \label{2.11}
\end{equation}%
In fact all average properties in the corresponding grand ensembles have a
similar relationship. For example, a property represented by the operator $%
\widehat{X}$ has the averages 
\begin{equation}
X_{e}\left( \beta ,V\mid \mu \right) \equiv \sum_{N=0}^{\infty
}Tr^{(N)}e^{\beta \Omega _{e}}e^{-\beta \left( \widehat{H}_{e}-\int d\mathbf{%
r}\mu (\mathbf{r})\widehat{n}(\mathbf{r})\right) }\widehat{X},  \label{2.11a}
\end{equation}%
\begin{equation}
X_{j}\left( \beta ,V\mid \mu \right) \equiv \sum_{N=0}^{\infty
}Tr^{(N)}e^{\beta \Omega _{j}}e^{-\beta \left( \widehat{H}_{j}-\int d\mathbf{%
r}\mu (\mathbf{r})\widehat{n}(\mathbf{r})\right) }\widehat{X},  \label{2.11c}
\end{equation}%
so with (\ref{2.5}) and (\ref{2.11}) they are related by%
\begin{equation}
X_{j}\left( \beta ,V\mid \mu \right) =X_{e}\left( \beta ,V\mid \mu
-v_{b}\right)  \label{2.11b}
\end{equation}

\subsection{Free energy density functionals}

The local number densities are%
\begin{equation}
n_{e}\left( \mathbf{r,}\beta ,V\mid \mu \right) =-\frac{\delta \beta \Omega
_{e}\left( \beta ,V\mid \mu \right) }{\delta \beta \mu (\mathbf{r})},\hspace{%
0.25in}n_{j}\left( \mathbf{r,}\beta ,V\mid \mu \right) =-\frac{\delta \beta
\Omega _{j}\left( \beta ,V\mid \mu \right) }{\delta \beta \mu (\mathbf{r})}.
\label{2.12}
\end{equation}%
The derivative of $\Omega _{j}\left( \beta ,V\mid \mu \right) $ is taken at
constant $n_{b}$. Using (\ref{2.11}) or (\ref{2.11b}) it is seen that the
number densities are related by%
\begin{equation}
n_{j}\left( \mathbf{r,}\beta ,V\mid \mu \right) =n_{e}\left( \mathbf{r,}%
\beta ,V\mid \mu -v_{b}\right) .  \label{2.13}
\end{equation}%
It can be shown that $\Omega _{e}\left( \beta ,V\mid \cdot \right) $ and $%
\Omega _{j}\left( \beta ,V\mid \cdot \right) $ (again with constant $n_{b}$)
are strictly concave functionals so that (\ref{2.12}) defines the one to one
invertible relationships $n_{e}\Longleftrightarrow \mu $ and $%
n_{j}\Longleftrightarrow \mu $. Consequently, a change of variables $\beta
,V,\mu \rightarrow \beta ,V,n_{e}$ for the inhomogeneous electron system and 
$\beta ,V,\mu \rightarrow \beta ,V,n_{j}$ for the inhomogeneous jellium are
possible. The corresponding Legendre transformations then define the free
energy functionals of these densities 
\begin{equation}
F_{e}\left( \beta ,V\mid n_{e}\right) =\Omega _{e}\left( \beta ,V\mid \mu
\right) +\int d\mathbf{r}\mu (\mathbf{r})n_{e}\left( \mathbf{r},\beta ,V\mid
\mu \right) ,  \label{2.14}
\end{equation}%
\begin{equation}
F_{j}\left( \beta ,V\mid n_{j}\right) =\Omega _{j}\left( \beta ,V\mid \mu
\right) +\int d\mathbf{r}\mu (\mathbf{r})n_{j}\left( \mathbf{r,}\beta ,V\mid
\mu \right) .  \label{2.15}
\end{equation}%
These are precisely the density functionals of DFT (e.g., defined by Mermin 
\cite{Mermin65}).

Their relationship follows using (\ref{2.5}) and (\ref{2.13}) 
\begin{eqnarray}
F_{j}\left( \beta ,V\mid n_{j}\right) &=&\beta E_{b}+\Omega _{e}\left( \beta
,V\mid \mu -v_{b}\right) +\int d\mathbf{r}\left( \mu (\mathbf{r})-v_{b}(%
\mathbf{r})\right) n_{e}\left( \mathbf{r,}\beta ,V\mid \mu -v_{b}\right) 
\notag \\
&&+\int d\mathbf{r}v_{b}(\mathbf{r})n_{e}\left( \mathbf{r,}\beta ,V\mid \mu
-v_{b}\right)  \notag \\
&=&F_{e}\left( \beta ,V\mid n_{e}\left( \mid \mu -v_{b}\right) \right) +\int
d\mathbf{r}v_{b}(\mathbf{r})n_{e}\left( \mathbf{r,}\beta ,V\mid \mu
-v_{b}\right) +\beta E_{b}  \label{2.16}
\end{eqnarray}%
and using (\ref{2.13}) again gives the desired result.%
\begin{equation}
F_{j}\left( \beta ,V\mid n_{j}\right) =F_{e}\left( \beta ,V\mid n_{j}\right)
+\int d\mathbf{r}v_{b}(\mathbf{r})n_{j}\left( \mathbf{r}\mid \mu \right)
+\beta E_{b}.  \label{2.17}
\end{equation}%
Since $\mu \left( \mathbf{r}\right) $ is arbitrary so also is $n_{j}\left( 
\mathbf{r}\mid \mu \right) $ and (\ref{2.17}) can be written more simply as%
\begin{equation}
F_{j}\left( \beta ,V\mid n\right) =F_{e}\left( \beta ,V\mid n\right) +\int d%
\mathbf{r}v_{b}(\mathbf{r})n\left( \mathbf{r}\right) +\beta E_{b}.
\label{2.18}
\end{equation}

\section{Exchange-correlation functional equivalence}

\label{sec3}Traditionally the free energy density functional is separated
into a non-interaction contribution, $F_{0}$, a mean-field Hartree
contribution, $F_{H}$, and the remaining exchange-correlation contribution, $%
F_{xc}$%
\begin{equation}
F=F_{0}+F_{H}+F_{xc}  \label{3.0}
\end{equation}%
Clearly $F_{0}$ is the same for the electron and jellium systems, as follows
from (\ref{2.18}) since $v_{b}(\mathbf{r})=0$ in this case). However, the
Hartree terms (defined as the average intrinsic internal energy with pair
correlation function equal to unity) are different%
\begin{equation}
F_{eH}\left( \beta ,V\mid n\right) =\frac{1}{2}e^{2}\int d\mathbf{r}d\mathbf{%
r}^{\prime }\frac{n(\mathbf{r})n(\mathbf{r}^{\prime })}{\left\vert \mathbf{r}%
-\mathbf{r}^{\prime }\right\vert },  \label{3.1}
\end{equation}%
\begin{equation}
F_{jH}\left( \beta ,V\mid n\right) =\frac{1}{2}e^{2}\int d\mathbf{r}d\mathbf{%
r}^{\prime }\frac{\left( n(\mathbf{r})-n_{b}\right) \left( n(\mathbf{r}%
^{\prime })-n_{b}\right) }{\left\vert \mathbf{r}-\mathbf{r}^{\prime
}\right\vert }.  \label{3.2}
\end{equation}%
In particular, their system size dependence is quite different. For example,
in the uniform density limit (assuming a spherical volume) $F_{eH}\left(
\beta ,V\mid n\right) \rightarrow C\left( n_{e}e\right) ^{2}V^{5/3}$ (with $%
C=\left( 4\pi \right) ^{2}\left( 1/15\right) \left( 3/\left( 4\pi \right)
\right) ^{5/3}\simeq \allowbreak 0.967$), whereas $F_{jH}\left( \beta ,V\mid
n\right) $ vanishes in this limit. This difference is the reason (along with
charge neutrality) why $F_{j}\left( \beta ,V\mid n\right) $ has a proper
thermodynamic limit \cite{Lieb75} while $F_{e}\left( \beta ,V\mid n\right) $
does not (e.g., it does not scale linearly with the volume).

At first sight this seems at odds with (\ref{2.18}) since the left side has
a thermodynamic limit whereas each term on the right separately does not.
However, the second and third terms cancel the singular volume dependence of 
$F_{eH}$ so that%
\begin{equation}
F_{eH}\left( \beta ,V\mid n\right) +\int d\mathbf{r}v_{b}(\mathbf{r})n\left( 
\mathbf{r}\right) +\beta E_{b}=F_{jH}\left( \beta ,V\mid n\right) .
\label{3.3}
\end{equation}%
This observation, together with the equavalence of the non-interacting
contributions, leads to the equivalence of the exchange - correlation
contributions for the electron gas and jellium%
\begin{equation}
F_{jxc}\left( \beta ,V\mid n\right) =F_{exc}\left( \beta ,V\mid n\right) .
\label{3.4}
\end{equation}%
Note that this equivalence applies for general inhomogeneous densities,
extending the familiar relationship for uniform systems.

These are two different inhomogeneous systems, yet their correlations are
the same for every admissable density. The functionals are each "universal"
in the sense that their forms are independent of the underlying external
potential. However, the potential associated with the chosen density of
their argument is different in each case. To see this consider the
functional derivative of (\ref{2.18}) with respect to the density

\begin{equation}
\frac{\delta F_{j}\left( \beta ,V\mid n\right) }{\delta n(\mathbf{r})}=\mu (%
\mathbf{r})=\frac{\delta F_{e}\left( \beta ,V\mid n\right) }{\delta n(%
\mathbf{r})}+v_{b}(\mathbf{r}).  \label{3.4a}
\end{equation}%
Hence the density $n$ is determined from the jellium functional by%
\begin{equation}
\frac{\delta F_{j}\left( \beta ,V\mid n\right) }{\delta n(\mathbf{r})}=\mu (%
\mathbf{r})=\mu -v_{ex}\left( \mathbf{r}\right) .  \label{3.4b}
\end{equation}%
Alternatively the same density is determined from the electron functional by%
\begin{equation}
\frac{\delta F_{e}\left( \beta ,V\mid n\right) }{\delta n(\mathbf{r})}=\mu (%
\mathbf{r})-v_{b}(\mathbf{r})=\mu -\left( v_{ex}(\mathbf{r})+v_{b}\left( 
\mathbf{r}\right) \right) .  \label{3.4c}
\end{equation}%
For jellium the potential is $v_{ex}\left( \mathbf{r}\right) $ while for the
electron gas it is $v_{ex}\left( \mathbf{r}\right) +v_{b}\left( \mathbf{r}%
\right) $. The two solutions to (\ref{3.4b}) and (\ref{3.4c}) have the same
relationship as expressed in (\ref{2.13}).

\section{Discussion}

Practical approximations for the electron gas exchange - correlation
functional are typically introduced at the level of its density, defined by%
\begin{equation}
F_{exc}\left( \beta ,V\mid n\right) =\int d\mathbf{r}f_{exc}\left( \mathbf{r}%
,\beta ,V\mid n\right) .  \label{4.1}
\end{equation}%
A formal functional expansion about the density at point $\mathbf{r}$ can be
performed%
\begin{equation*}
f_{exc}\left( \mathbf{r},\beta ,V\mid n\right) =f_{exc}\left( \mathbf{r}%
,\beta ,V\mid n\right) \mid _{n\left( \mathbf{r}\right) }+\int d\mathbf{r}%
^{\prime }\frac{\delta f_{exc}\left( \mathbf{r},\beta ,V\mid n\right) }{%
\delta n\left( \mathbf{r}^{\prime }\right) }\mid _{n\left( \mathbf{r}\right)
}\left( n\left( \mathbf{r}^{\prime }\right) -n\left( \mathbf{r}\right)
\right)
\end{equation*}%
\begin{equation}
+\int d\mathbf{r}^{\prime }d\mathbf{r}^{\prime \prime }\frac{\delta
f_{exc}\left( \mathbf{r},\beta ,V\mid n\right) }{\delta n\left( \mathbf{r}%
^{\prime }\right) \delta n\left( \mathbf{r}^{\prime \prime }\right) }\mid
_{n\left( \mathbf{r}\right) }\left( n\left( \mathbf{r}^{\prime }\right)
-n\left( \mathbf{r}\right) \right) \left( n\left( \mathbf{r}^{\prime \prime
}\right) -n\left( \mathbf{r}\right) \right) +..  \label{4.2}
\end{equation}%
The coefficients are evaluated at the "uniform density" $n\left( \mathbf{r}%
\right) $, i.e. all functional density dependence is evaluated at the same
value. Consequently, the lead term is just the uniform electron gas exchange
- correlation free energy per unit volume 
\begin{equation}
f_{exc}\left( \mathbf{r},\beta ,V\mid n\right) \mid _{n\left( \mathbf{r}%
\right) }=\frac{1}{V}F_{exc}\left( \beta ,V,n_{e}\right) \mid
_{n_{e}=n\left( \mathbf{r}\right) }.  \label{4.3}
\end{equation}%
This is known as the "local density approximation. Similarly the subsequent
terms in (\ref{4.2}) are the response functions for the uniform electron
gas. From the above analysis all of these can now\ be identifed with those
for jellium, which has been studied extensively. As noted in the
Introduction, an accurate analytic fit for $F_{jxc}\left( \beta
,V,n_{e}\right) $ is now available across the entire $\beta ,n_{e}$ plane 
\cite{KSDT15}, so the local density approximation is known explicitly.
Similarly, the first few response functions for jellium are known as well 
\cite{Vignale05}.

More generally approximations for exchange - correlations away from the
uniform limit (e.g., generalized gradient approximations) can be addressed
for \textit{inhomogeneous} jellium as well. While this is a difficult
problem it is placed in a more controlled thermodynamic context due to
charge neutrality and extensivity.

\section{Acknowledgements}

\label{sec5}This research was supported by US DOE Grant DE-SC0002139.

\bigskip

\end{document}